\documentclass[aps,prc,preprint,amsmath,amssymb,showpacs,preprintnumbers,superscriptaddress]{revtex4-1}
\usepackage{CJK}
\usepackage{graphicx}
\usepackage{dcolumn}
\usepackage{bm}
\usepackage{color}
\usepackage{multirow}
\usepackage{amsmath}
\everymath{\displaystyle}
\allowdisplaybreaks[4]
\renewcommand\arraystretch{1.5}
\usepackage{hyperref}
\begin{document}

\title{Selection rules of electromagnetic transitions for chirality-parity violation in atomic nuclei}
\author{Y. Y. Wang}
\affiliation{State Key Laboratory of Nuclear Physics and Technology, School of Physics, Peking University, Beijing 100871, China}
\author{X. H. Wu}
\affiliation{State Key Laboratory of Nuclear Physics and Technology, School of Physics, Peking University, Beijing 100871, China}
\author{S. Q. Zhang}\email{sqzhang@pku.edu.cn}
\affiliation{State Key Laboratory of Nuclear Physics and Technology, School of Physics, Peking University, Beijing 100871, China}
\author{P. W. Zhao}
\affiliation{State Key Laboratory of Nuclear Physics and Technology, School of Physics, Peking University, Beijing 100871, China}
\author{J. Meng }\email{mengj@pku.edu.cn}
\affiliation{State Key Laboratory of Nuclear Physics and Technology, School of Physics, Peking University, Beijing 100871, China}
\affiliation{School of Physics and Nuclear Energy Engineering, Beihang University, Beijing 100191, China}
\affiliation{Yukawa Institute for Theoretical Physics, Kyoto University, Kyoto 606-8502, Japan}

\date{\today}
\begin{abstract}
The nuclear Chirality-Parity (ChP) violation, a simultaneous breaking of chiral and reflection symmetries in the intrinsic frame, is investigated with a reflection-asymmetric triaxial particle rotor model.
A new symmetry for an ideal ChP violation system is found and the corresponding selection rules of the electromagnetic transitions are derived.
The fingerprints for the ChP violation including the nearly degenerate quartet bands and the selection rules of the electromagnetic transitions are provided.
These fingerprints are examined for ChP quartet bands by taking a two-$j$ shell $h_{11/2}$ and $d_{5/2}$ with typical energy spacing for $A=$ 130 nuclei.
\end{abstract}
\maketitle
\date{today}

As a microscopic quantum many-body system, the atomic nucleus carries a wealth of information on fundamental symmetries and symmetry breakings which are usually manifested in energy spectra and electromagnetic transitions.
For example, the spherical symmetry in atomic nuclei results in equally spaced vibration spectra, and the deviation from spherical symmetry may lead to a series of rotor-like sequences~\cite{bohr1975nuclear}.

The reflection symmetry and chiral symmetry have broad interests in mathematics, physics, chemistry and biology.
In nuclear physics, the reflection symmetry~\cite{Bohr1956} and chiral symmetry~\cite{Frauendorf1997Tilted} have been at the frontiers, and continue to be hot topics over the past decades, see e.g., reviews~\cite{Butler1996Rev.Mod.Phys.349,Butler_2016,Frauendorf2001Spontaneous,Meng2010Open,Meng2016Nuclear,
Raduta2016Prog.Part.Nucl.Phys.241,Frauendorf2018Phys.Scripta43003,Xiong2019Atom.DataNucl.DataTabl.193}.

In pear-shaped nuclei, the spontaneous breaking of reflection symmetry occurs and manifests itself by the occurrence of the interleaved positive and negative parity bands or a pair of parity partner bands~\cite{Butler1996Rev.Mod.Phys.349,Butler_2016}.
The pear-shaped nuclei can provide a unique probe to test the charge-parity (CP) symmetry violation beyond the standard model~\cite{gaffney2013studies}, and have attracted a lot of attentions in both nuclear physics and particle physics.

In triaxially deformed nuclei, the spontaneous breaking of chiral symmetry may occur and manifests itself by the appearance of chiral doublet bands, a pair of nearly degenerate $\Delta I=1\hbar$ bands with the same parity~\cite{Frauendorf1997Tilted}.
The coexistence of two or more chiral doublet bands in a single nucleus, i.e., multiple chiral doublets (M$\chi$D) predicted by the microscopic covariant density functional theory~\cite{Meng2006Possible,Zhao2017Phys.Lett.B1}, has been observed in $^{133}$Ce~\cite{Ayangeakaa2013Evidence}, $^{103}$Rh \cite{Kuti2014Multiple}, $^{78}$Br~\cite{Liu2016Phys.Rev.Lett.112501},
$^{136}$Nd~\cite{Petrache2018Phys.Rev.C41304}, $^{195}$Tl~\cite{Roy2018Phys.Lett.768}, and $^{135}$Nd~\cite{Zhu2003A,Mukhopadhyay2007From,PhysRevC.100.024314}.

\begin{figure}[!htbp]
  \centerline{
  \includegraphics[width=0.6\textwidth]{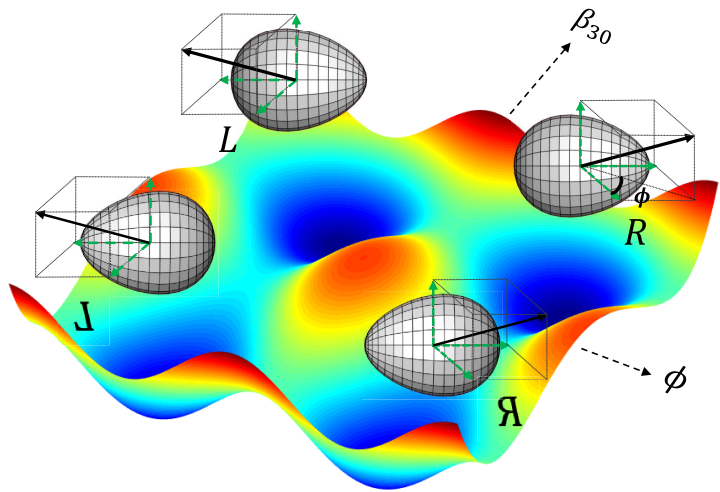}}
  \caption{A schematic potential energy surface in the intrinsic $(\beta_{30},\phi)$ plane for a ChP violation system, with $\beta_{30}$ the octupole deformation parameter and $\phi$ the azimuthal orientation angle of the total angular momentum.
   The sign of $\beta_{30}$ stands for the orientation of the nuclear distribution parallel or antiparallel with the intrinsic axis. The sign of $\phi$ stands for the right- and left-handed system.}
  \label{fig1}
\end{figure}

The observation in $^{78}$Br, two pairs of chiral doublet bands with opposite parity connected with strong electric dipole ($E1$) transitions, provides the first evidence of M$\chi$D bands in octupole soft nuclei~\cite{Liu2016Phys.Rev.Lett.112501}.
This observation indicates that nuclear chirality can be robust against the octupole correlations, which together with the scenario in Ref.~\cite{Frauendorf2001Spontaneous} encourages the exploration of the simultaneous chiral and reflection symmetry breaking in a reflection-asymmetric triaxial nucleus.

A schematic potential energy surface with simultaneous chiral and reflection symmetry breaking in the intrinsic $(\beta_{30},\phi)$ plane is given in Fig.~\ref{fig1}, with $\beta_{30}$ the octupole deformation parameter and $\phi$ the azimuthal orientation angle of the total angular momentum.
There are four minima corresponding to the orientation of the nuclear distribution parallel or antiparallel with the intrinsic axis, and of the angular momentum in the intrinsic frame.
The symmetry restoration in the laboratory frame results in both chiral and parity splittings which give rise to four nearly degenerate states, i.e., chirality-parity (ChP) quartet states.
The rotational excitation on the ChP quartet states will generate four nearly degenerate rotational bands, i.e., the so-called ChP quartet bands, as expected in Refs.~\cite{Frauendorf2001Spontaneous,
Liu2016Phys.Rev.Lett.112501}.

In this paper, the nuclear ChP violation, a simultaneous breaking of chiral and reflection symmetries in the intrinsic frame, is investigated with a reflection-asymmetric triaxial particle rotor model (RAT-PRM).
The energy spectra and electromagnetic transitions will be investigated, and the possible
fingerprints for ChP quartet bands will be explored.

The ideal chiral system prefers a maximum triaxial deformation and a pure particle-hole configuration~\cite{Frauendorf1997Tilted}.
The ideal reflection-asymmetric system requires a static octupole deformation resulting from strong octupole correlations.
In order to describe the ideal ChP violation system, a RAT-PRM  is developed for a core with triaxiality $\gamma=90^\circ$ coupled with one particle and one hole in a two-$j$ shell with its orbital and total angular momenta differing by 3$\hbar$, i.e., $\Delta l=\Delta j=3\hbar$.

The Hamiltonian of the core is written as
\begin{align}
  \hat{H}_{\rm core} = \frac{1}{2\mathcal{J}_0} \Bigg[ \hat{R}_3^2 +4(\hat{R}_1^2+\hat{R}_2^2)\Bigg] + \frac{1}{2}E(0^-)(1-\hat{P}_{c}), \label{eq-hcore}
\end{align}
where the moments of inertia for irrotational flow $\mathcal{J}_k=\mathcal{J}_0\sin^2(\gamma-2k\pi/3)$ are adopted,
and $\hat{R}_k$ with $k=1,2,3$ represent the core angular momentum along the long, short, and intermediate axes, respectively.
The second term describes the parity splitting of the reflection-asymmetric core, with $\hat{P}_{c}$ the core parity operator and $E(0^-)$ the splitting parameter~\cite{WANG2019454,wang2020}.

The intrinsic Hamiltonian for the particle can be written as
\begin{align}
 \hat{H}_{\rm s.p.} & = \hat{h}_{lj}
                      -\hbar\omega_0r^2\left[
                      \frac{\beta_{22}}{\sqrt{2}}(Y_{22}+Y_{2-2})+\beta_{30}Y_{30}\right],
                      \label{eq-vp}
\end{align}
where $\hat{h}_{lj}$ is the spherical single-particle Hamiltonian with $l$ the orbital angular momentum and $j$ the total angular momentum, $\hbar\omega_0=41A^{-1/3}$ MeV, and $\beta_{22}$ and $\beta_{30}$ represent the quadrupole and octupole deformation parameters, respectively.
The intrinsic Hamiltonian for the hole can be obtained by changing the sign of $\hat{H}_{\rm s.p.}$\cite{Frauendorf1996Interpretation}.

The core Hamiltonian in Eq.~(\ref{eq-hcore}) has good parity $P_c$ and $D_2$ symmetry, i.e., the invariance with respect to rotations by $\pi$ about each of the three principal axes.

The intrinsic Hamiltonian in Eq.~(\ref{eq-vp}) has $V_4$ symmetry, i.e., the invariance with respect to the reversion of the intrinsic 1 and 2 axes, while the octupole deformation breaks the intrinsic space-reflection symmetry and $D_2$ symmetry.

The RAT-PRM is constructed in the laboratory frame and its Hamiltonian consisting of the core and intrinsic parts has good total parity $P$, good angular momentum $I$, and $V_4$ symmetry.
Moreover, if the two-$j$ shell for the particle and hole is the same and the $lj$ dependence of the corresponding $\hat{h}_{lj}$ in Eq.~(\ref{eq-vp}) is neglected, the total Hamiltonian is invariant under the operation $\hat{\mathcal{A}}=\hat{\mathcal{R}}_3(\frac{\pi}{2})\hat{C}\hat{\pi}$, with $\hat{\mathcal{R}}_3(\frac{\pi}{2})$ the rotation by $\frac{\pi}{2}$ around 3-axis, $\hat{C}$ the exchange of particle and hole, and $\hat{\pi}$ the intrinsic space-reflection.
For the reflection-symmetric system, the intrinsic Hamiltonian has good parity, the corresponding operator $\hat{\mathcal{A}}$ becomes $\hat{\mathcal{R}}_3(\frac{\pi}{2})\hat{C}$, which has been discussed in Ref.~\cite{Koike2004Chiral} together with the characteristic of the electromagnetic transitions for the chiral doublet bands.

The operation $\hat{\mathcal{A}}$ plays a role in the exchange of the right-handed system with left-handed system by the exchange of the particle and hole.
The corresponding quantum number $\mathcal{A}$ is named as ``\emph{chiture}'', in analogy to the $\emph{signature}$ quantum number for the rotational operation $\hat{\mathcal{R}}(\pi)$.
Similar to the $\emph{simplex}$ operator $\hat{\mathcal{S}}=\hat{\mathcal{R}}(\pi)\hat{P}$, a new operator $\hat{\mathcal{B}}=\hat{\mathcal{A}}\hat{P}$ is introduced and the corresponding quantum number $\mathcal{B}$ is named as ``\emph{chiplex}''.

For a given angular momentum $I$, the eigenstates of the RAT-PRM Hamiltonian can be characterized by the total parity $P$ and the chiplex $\mathcal{B}$, as shown in Table~\ref{table1}.
The corresponding chiture $\mathcal{A}$ is given in the third column.
The $V_4$ symmetry of the total Hamiltonian requires the projection of the core angular momentum $R_3$ to be even integers.
For $\mathcal{A}=+1$, $C\pi=$ $+1$ or $-1$, and the corresponding $R_3$ are $0, \pm 4, \pm8, \cdots$ or $\pm 2, \pm6, \cdots$.
For details, see Table~\ref{table1}.
The states with same parity $P$ but different chiture $\mathcal{A}$ constitute the chiral doublets, and the states with same chiture $\mathcal{A}$ but different parity $P$ constitute the parity doublets.

\renewcommand\arraystretch{0.94}
\begin{table}[h!]
\centering
\caption{Parity $P$ and chiplex $\mathcal{B}$ for the eigenstates of the RAT-PRM Hamiltonian.
The chiture $\mathcal{A}$ as well as possible $R_3$ and $C\pi$ values under $V_4$ symmetry is also listed.}
\begin{tabular}{c|c|c|c|c}
\hline
\hline
\multirow{1}*{~~~~$P$~~~~}  & \multirow{1}*{~~~~$\mathcal{B}$~~~~} & \multirow{1}*{~~~~$\mathcal{A}$~~~~} & \multirow{1}*{~~~~$R_3$~~~~}  & \multirow{1}*{~~~~$C\pi$~~~~}  \\
\hline
\multirow{4}*{~~~$+1$~~~}  & \multirow{2}*{~~~$+1$~~~} & \multirow{2}*{~~~$+1$~~~}  & $0,\pm4,\pm8\cdots$ & $+1$ \\
 \cline{4-5}
 & & & $\pm2,\pm6 \cdots$ & $-1$  \\
 \cline{2-5}
 &\multirow{2}*{~~~$-1$~~~} & \multirow{2}*{~~~$-1$~~~} & $0,\pm4,\pm8\cdots$ & $-1$ \\
 \cline{4-5}
  & & & $\pm2,\pm6 \cdots$ & $+1$  \\
 \cline{2-5}
\hline
\multirow{4}*{~~~$-1$~~~}  & \multirow{2}*{~~~$+1$~~~} & \multirow{2}*{~~~$-1$~~~}  & $0,\pm4,\pm8\cdots$ & $-1$ \\
 \cline{4-5}
 & & & $\pm2,\pm6 \cdots$ & $+1$  \\
 \cline{2-5}
 &\multirow{2}*{~~~$-1$~~~} & \multirow{2}*{~~~$+1$~~~} & $0,\pm4,\pm8\cdots$ & $+1$ \\
 \cline{4-5}
  & & & $\pm2,\pm6 \cdots$ & $-1$  \\
 \cline{2-5}
\hline
\hline
\end{tabular}\label{table1}
\end{table}

The eigenstates of the RAT-PRM Hamiltonian with different angular momenta are connected by electromagnetic transitions, including $E2$ and $M1$ transitions between same parity states, and $E1$ and $E3$ transitions between opposite parity states.

The $E2$ transitions can be calculated from the operator, $\hat{\mathcal{M}}(E2)=\sum_{\mu=0,\pm2}\hat{q}_{2\mu}^{(c)}$.
As the intrinsic part of $\hat{\mathcal{M}}(E2)$ is neglected, it connects states with the same $C$ and $\pi$, i.e., $\Delta C=0$ and $\Delta \pi=0$.
For $\gamma=90^\circ$, the operator $\hat{q}_{20}^{(c)}\propto \beta_2\cos\gamma$ does not contribute and thus the matrix elements between states with same $R_3$ vanish.
Therefore, the $E2$ matrix elements occur only between states with different $\mathcal{A}$ and same $C\pi$, or in other words with different $\mathcal{B}$ and same $C\pi$.

The $M1$ transitions can be calculated from the operator, $$\hat{\mathcal{M}}(M1)=\sum_{\mu=0,\pm1}\sqrt{\frac{3}{4\pi}}\frac{e\hbar}{2Mc}[(g_p-g_R)\hat{j}^p_{1\mu}+(g_n-g_R)\hat{j}^n_{1\mu}].$$
If the effective gyromagnetic factor $g_{p(n)}-g_R$ is set equal to 1 for proton and -1 for neutron~\cite{Frauendorf1997Tilted}, the $M1$ transitions are contributed from the matrix elements between states with the same $\pi$ and different $C$. 
Furthermore, the $M1$ operator changes $R_3$ by $\Delta R_3=0,\pm1$.
Therefore, the $M1$ matrix elements occur only between states with different $\mathcal{A}$ and $C\pi$, or in other words with different $\mathcal{B}$ and $C\pi$.

The $E3$ transitions can be calculated from the operator, $\hat{\mathcal{M}}(E3)=\hat{q}_{30}^{(c)}$.
As the intrinsic part is neglected, one has $\Delta C=0$ and $\Delta \pi=0$.
For the core, with the operator $\hat{q}_{30}^{(c)}\propto\beta_{30}$, the contributions are from the matrix elements between states with same $R_3$.
Therefore, the $E3$ matrix elements contribute only between states with same $\mathcal{A}$ and $C\pi$, or in other words with different $\mathcal{B}$ and same $C\pi$.

There is no $E1$ transitions here.
For $\gamma=90^\circ$, the collective $E1$ operator $\hat{q}_{10}^{(c)}\propto \beta_2\beta_{30}\cos\gamma$ vanishes~\cite{WANG2019454}.
The intrinsic part does not contribute because the model space is limited to a two-$j$ shell with $\Delta l=\Delta j=3\hbar$.

\begin{figure}[!htbp]
  \centerline{
  \includegraphics[width=0.75\textwidth]{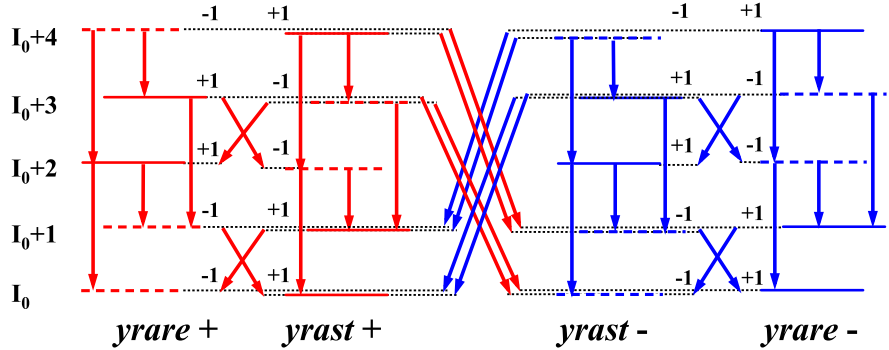}}
  \caption{
  The ChP quartet bands organized by two pairs of chiral doublet bands with positive parity denoted by yrast$+$ and yrare$+$, and negative parity denoted by yrast$-$ and yrare$-$.
  States with $\mathcal{B}=+1 (-1)$ are denoted by solid (dashed) lines.
  Allowed $E2$, $M1$, and $E3$ transitions are denoted by arrows.}
  \label{fig2}
\end{figure}

Starting from the yrast and yrare chiral doublets at $I_0$ for the positive and negative parity, with the $E2$ transitions between $\Delta I=2\hbar$ allowed, two pairs of chiral doublet bands are organized.
They constitute the ChP quartet bands, as shown in Fig.~\ref{fig2}.
The sign of the chiplex $\mathcal{B}$ between chiral partners is opposite, which is also true between parity partners.
As the $E2$ transitions occur only between states with different $\mathcal{B}$, $\mathcal{B}$ changes the sign every $2\hbar$ in a band and there is no interband $E2$ transitions.

In Fig.~\ref{fig2}, between states with different chiplex $\mathcal{B}$, the allowed $M1$ transitions with $\Delta I=1\hbar$, and the allowed $E3$ transitions with $\Delta I=3\hbar$ are also shown.
The intraband and interband $B(M1)$ for chiral doublet bands exhibit staggering behavior.
The alternating interband $E3$ transitions with spin between yrast$+$ and yrast$-$ might serve as a fingerprint for ChP quartet bands.
Similar alternating interband $E3$ transitions between yrast$+$ and yrare$-$, between yrare$+$ and yrast$-$, and between yrare$+$ and yrare$-$, are also expected.

The allowed and forbidden $E2$, $M1$, and $E3$ transitions for ChP quartet bands by chiplex $\mathcal{B}$ and parity $P$ discussed above are summarized in Table~\ref{table2}.

\renewcommand\arraystretch{0.8}
\begin{table}[h!]
\centering
\caption{The electromagnetic transitions for the ChP quartet bands.
The allowed transitions are denoted by ``$\surd$'', the chiplex forbidden transitions are denoted by ``$\times$'', and the parity forbidden transitions are denoted by ``$-$''.}
\begin{tabular}{c|c|c|c|c|c}
\hline
\hline
\multicolumn{1}{c|}{~~~~~~~~}  & \multicolumn{1}{c|}{~~~$E2$~~~} & \multicolumn{2}{c|}{~~$M1$~~} & \multicolumn{2}{c}{$~E3$~}  \\
\hline
\multicolumn{1}{c|}{~~Intra-band~~}  & $I\rightarrow I-2$ & $I\rightarrow I-1$ & $I+1\rightarrow I$& $I\rightarrow I-3$ & $I+1\rightarrow I-2$ \\
\hline
 yrast$+\rightarrow$ yrast$+$ & \multirow{4}*{~~$\surd$~~} & \multirow{4}*{~~$\surd$~~}  & \multirow{4}*{~~~$\times$~~~}  & \multirow{4}*{~~~$-$~~~} & \multirow{4}*{~~~$-$~~~} \\
 yrare$+\rightarrow$ yrare$+$ & & & & & \\
 yrast$-\rightarrow$ yrast$-$ & & & & & \\
 yrare$-\rightarrow$ yrare$-$ & & & & & \\
\hline
\multicolumn{1}{c|}{~~Inter-band~~}  & ~~~$I\rightarrow I-2$~~~ & $I\rightarrow I-1$ & ~~~$I+1\rightarrow I$~~~  & ~$I\rightarrow I-3$~ & $I+1\rightarrow I-2$ \\
\hline
 yrast$+\leftrightarrow$ yrare$+$ & \multirow{2}*{~~~$\times$~~~} & \multirow{2}*{~~~$\times$~~~}  & \multirow{2}*{~~~$\surd$~~~} &  \multirow{2}*{~~~$-$~~~} & \multirow{2}*{~~~$-$~~~} \\
 yrast$-\leftrightarrow$ yrare$-$ & & & & & \\
\cline{1-6}
 yrast$+\leftrightarrow$ yrast$-$ & \multirow{2}*{~~~$-$~~~} & \multirow{2}*{~~~$-$~~~}  & \multirow{2}*{~~~$-$~~~} &  \multirow{2}*{~~~$\surd$~~~} & \multirow{2}*{~~~$\times$~~~}  \\
 yrare$+\leftrightarrow$ yrare$+$ & & & & & \\
\cline{1-6}
 yrast$+\leftrightarrow$ yrare$-$ &\multirow{2}*{~~~$-$~~~} & \multirow{2}*{~~~$-$~~~}  & \multirow{2}*{~~~$-$~~~} &  \multirow{2}*{~~~$\times$~~~} & \multirow{2}*{~~~$\surd$~~~}  \\
 yrare$+\leftrightarrow$ yrast$-$ & & & & & \\
\hline
\hline
\end{tabular}\label{table2}
\end{table}

The schematic energy spectra and electromagnetic transitions for the ChP quartet bands in Fig.~\ref{fig2} are based on the symmetry analysis, which is for the ideal case, i.e., $\gamma=90^\circ$, $E(0^-)=0$ MeV, the intrinsic configuration includes one particle and one hole, and the two-$j$ shell is degenerate.

It is interesting to examine the ChP quartet bands, including their energy spectra, electromagnetic transitions and chiral geometry, if the restriction for a constant $\hat{h}_{lj}$ in Eq.~(\ref{eq-vp}) is released.

For nuclei in $A=$ 130 mass region, the two-$j$ shell with $\Delta l=\Delta j=3\hbar$ near the Fermi surface is $h_{11/2}$ and $d_{5/2}$, and their energy difference is taken as 0.89 MeV~\cite{nilsson1969nuclear}.
The other parameters adopted in RAT-PRM include the moment of inertia $\mathcal{J}_0=18\hbar^2/$MeV and the parity splitting parameter $E(0^-)=0$ MeV for the core, and the deformation parameters $\beta_{22}=0.3$ and $\beta_{30}=0.1$ for the intrinsic Hamiltonian.

\begin{figure}[!htbp]
  \centerline{
  \includegraphics[width=0.85\textwidth]{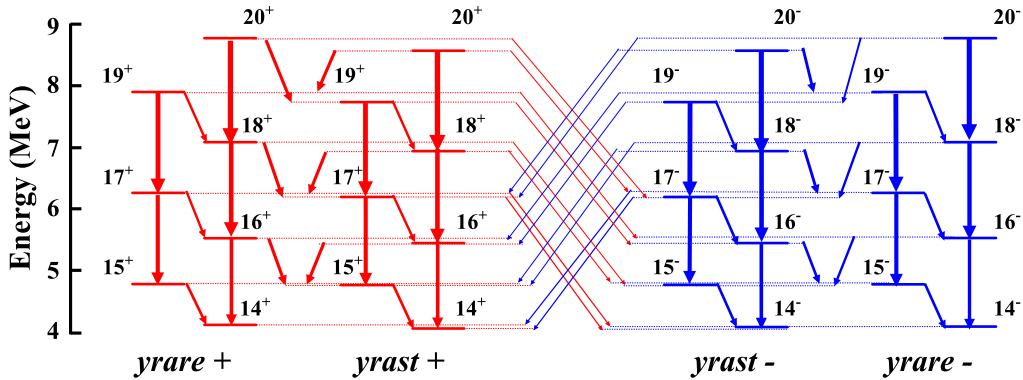}}
  \caption{The ChP quartet bands in RAT-PRM respectively denoted by yrast$+$, yrare$+$, yrast$-$ and yrare$-$.
  The transitions $E2$, $M1$, and $E3$ are denoted by arrows and the thicknesses of the arrows stand for their strength. For $M1$ transitions, $g_{p(n)}-g_R=1(-1)$ is adopted.  
 }
  \label{fig3}
\end{figure}

Diagonalizing the RAT-PRM Hamiltonian, the energy spectra can be obtained.
In Fig.~\ref{fig3}, the calculated ChP quartet bands are given for $14\hbar \leq I \leq 20\hbar$ which corresponds to the region of the static chirality.
For both the positive and negative parity, the splitting between the chiral doublet bands is less than 200 keV.
The splitting between the parity doublet bands is rather small (less than 20 keV) since the core-parity splitting $E(0^-)=0$ MeV.

The electromagnetic transitions follow the features in Fig.~\ref{fig2}.
The intraband $B(E2)$ is about one order of magnitude stronger than the interband one.
Both the intraband $B(M1)$ and the interband $B(M1)$ exhibit staggering behavior.
The alternating interband $E3$ transitions with spin between yrast$+$ and yrast$-$ appear, in consistent with the fingerprint for ChP quartet bands.
Similarities for $B(E3)$ between yrast$+$ and yrare$-$, between yrare$+$ and yrast$-$, and between yrare$+$ and yrare$-$, are also shown.

\begin{figure}[!htbp]
  \centerline{
  \includegraphics[width=0.65\textwidth]{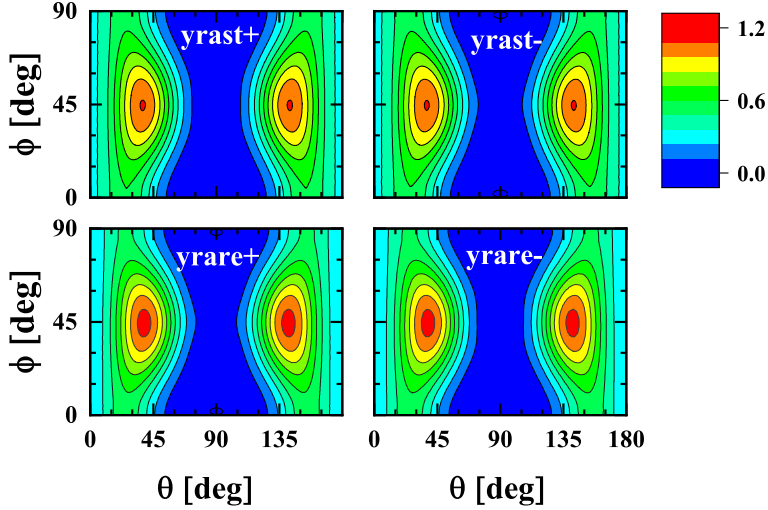}}
  \caption{The \emph{azimuthal plot}, i.e., the probability profile for the orientation of the angular momentum on the $(\theta,\phi)$ plane, for ChP quartet states at $I=16\hbar$ in RAT-PRM.
  }
  \label{fig4}
\end{figure}

The chiral geometry for the chiral doublet bands is usually analyzed by the composition of the angular momentum~\cite{Zhang2007Phys.Rev.C44307,Qi2009Chirality}, the $K-$plot~\cite{Qi2009Chirality}, the effective angles~\cite{Wang2007Phys.Rev.C24309,Chen2018Phys.Rev.31303}, as well as the \emph{azimuthal plot}~\cite{Chen2016Two,Chen2017Chiral,Chen2018Phys.Rev.31303}.

The \emph{azimuthal plot} provides the probability profile for the orientation of the angular momentum on the $(\theta,\phi)$ plane, with $\theta$ the polar angle between the angular momentum and the intrinsic 3-axis and $\phi$ the azimuthal angle between the projection of the angular momentum on the intrinsic 1-2 plane and the 1-axis.
For $\gamma=90^\circ$, the intrinsic 1, 2, and 3 axes are the long ($l$), short ($s$), and intermediate ($i$) axes, respectively.
Accordingly,  $\theta$ plays the role of $\phi$ in Fig.~\ref{fig1}.

In Fig.~\ref{fig4}, the \emph{azimuthal plot} for ChP quartet states at $I=16\hbar$ in RAT-PRM is shown.
The probability profiles for the ChP quartet states peak at ($\phi\sim 45^\circ$, $\theta \sim35^\circ$) and ($\phi\sim 45^\circ$, $\theta \sim145^\circ$), demonstrating the static chirality for the chiral doublets and the near identical distribution for parity doublets.

In summary, the nuclear ChP violation, a simultaneous breaking of chiral and reflection symmetries in the intrinsic frame, is investigated with a RAT-PRM.
The ideal ChP violation system corresponds a maximum triaxial deformation, a pure particle-hole configuration, and a static octupole deformation.
The RAT-PRM Hamiltonian for the ideal ChP violation system has good total parity $P$, good angular momentum $I$, and $V_4$ symmetry.
A new symmetry, chiture $\hat{\mathcal{A}}$ or chiplex $\hat{\mathcal{B}}=\hat{\mathcal{A}}\hat{P}$, is derived for $\gamma=90^\circ$.
The eigenstates of the RAT-PRM Hamiltonian can be characterized by the total parity $P$ and the chiplex $\mathcal{B}$.

The selection rules of the electromagnetic transitions for ChP violation are revealed, and $E2$, $M1$, and $E3$ transitions are found to link the states with different $\mathcal{B}$ only.
Starting from the yrast and yrare chiral doublets at $I_0$ for the positive parity and negative parity, the ChP quartet bands are constructed with the $E2$ transitions between $\Delta I=2\hbar $ allowed.  
Both the intraband $B(M1)$ and interband $B(M1)$ exhibit staggering behavior.
The interband $E3$ transitions alternate with spin.

The fingerprints for ChP quartet bands including the nearly degeneracy in energy and the selection rules of electromagnetic transitions are examined by taking a two-$j$ shell $h_{11/2}$ and $d_{5/2}$ with typical energy spacing for $A=$ 130 nuclei.
The static chirality for the chiral doublets and the near identical distribution for parity doublets are demonstrated by the \emph{azimuthal plot}.

\begin{acknowledgments}
This work was partly supported by the National Natural Science Foundation of China (Grants No. 11875075, No. 11935003, No. 11975031, and No. 11621131001), the National Key R\&D Program of China (Contracts No. 2018YFA0404400 and No. 2017YFE0116700), and the State Key Laboratory of Nuclear Physics and Technology, Peking University (No. NPT2020ZZ01).
\end{acknowledgments}

%

\begin{thebibliography}{33}%
\makeatletter
\providecommand \@ifxundefined [1]{%
 \@ifx{#1\undefined}
}%
\providecommand \@ifnum [1]{%
 \ifnum #1\expandafter \@firstoftwo
 \else \expandafter \@secondoftwo
 \fi
}%
\providecommand \@ifx [1]{%
 \ifx #1\expandafter \@firstoftwo
 \else \expandafter \@secondoftwo
 \fi
}%
\providecommand \natexlab [1]{#1}%
\providecommand \enquote  [1]{``#1''}%
\providecommand \bibnamefont  [1]{#1}%
\providecommand \bibfnamefont [1]{#1}%
\providecommand \citenamefont [1]{#1}%
\providecommand \href@noop [0]{\@secondoftwo}%
\providecommand \href [0]{\begingroup \@sanitize@url \@href}%
\providecommand \@href[1]{\@@startlink{#1}\@@href}%
\providecommand \@@href[1]{\endgroup#1\@@endlink}%
\providecommand \@sanitize@url [0]{\catcode `\\12\catcode `\$12\catcode
  `\&12\catcode `\#12\catcode `\^12\catcode `\_12\catcode `\%12\relax}%
\providecommand \@@startlink[1]{}%
\providecommand \@@endlink[0]{}%
\providecommand \url  [0]{\begingroup\@sanitize@url \@url }%
\providecommand \@url [1]{\endgroup\@href {#1}{\urlprefix }}%
\providecommand \urlprefix  [0]{URL }%
\providecommand \Eprint [0]{\href }%
\providecommand \doibase [0]{http://dx.doi.org/}%
\providecommand \selectlanguage [0]{\@gobble}%
\providecommand \bibinfo  [0]{\@secondoftwo}%
\providecommand \bibfield  [0]{\@secondoftwo}%
\providecommand \translation [1]{[#1]}%
\providecommand \BibitemOpen [0]{}%
\providecommand \bibitemStop [0]{}%
\providecommand \bibitemNoStop [0]{.\EOS\space}%
\providecommand \EOS [0]{\spacefactor3000\relax}%
\providecommand \BibitemShut  [1]{\csname bibitem#1\endcsname}%
\let\auto@bib@innerbib\@empty
\bibitem [{\citenamefont {Bohr}\ and\ \citenamefont
  {Mottelson}()}]{bohr1975nuclear}%
  \BibitemOpen
  \bibfield  {author} {\bibinfo {author} {\bibfnamefont {A.}~\bibnamefont
  {Bohr}}\ and\ \bibinfo {author} {\bibfnamefont {B.~R.}\ \bibnamefont
  {Mottelson}},\ }\href@noop {} {\emph {\bibinfo {title} {Nuclear
  Structure}}},\ Vol.~\bibinfo {volume} {II}\ (\bibinfo  {publisher} {Benjamin,
  New York, 1975})\BibitemShut {NoStop}%
\bibitem [{\citenamefont {Bohr}(1956)}]{Bohr1956}%
  \BibitemOpen
  \bibfield  {author} {\bibinfo {author} {\bibfnamefont {A.}~\bibnamefont
  {Bohr}},\ }\href {\doibase 10.1007/BF02744334} {\bibfield  {journal}
  {\bibinfo  {journal} {Il Nuovo Cimento}\ }\textbf {\bibinfo {volume} {4}},\
  \bibinfo {pages} {1091} (\bibinfo {year} {1956})}\BibitemShut {NoStop}%
\bibitem [{\citenamefont {Frauendorf}\ and\ \citenamefont
  {Meng}(1997)}]{Frauendorf1997Tilted}%
  \BibitemOpen
  \bibfield  {author} {\bibinfo {author} {\bibfnamefont {S.}~\bibnamefont
  {Frauendorf}}\ and\ \bibinfo {author} {\bibfnamefont {J.}~\bibnamefont
  {Meng}},\ }\href
  {http://www.sciencedirect.com/science/article/pii/S0375947497000043}
  {\bibfield  {journal} {\bibinfo  {journal} {Nucl. Phys. A}\ }\textbf
  {\bibinfo {volume} {617}},\ \bibinfo {pages} {131} (\bibinfo {year}
  {1997})}\BibitemShut {NoStop}%
\bibitem [{\citenamefont {Butler}\ and\ \citenamefont
  {Nazarewicz}(1996)}]{Butler1996Rev.Mod.Phys.349}%
  \BibitemOpen
  \bibfield  {author} {\bibinfo {author} {\bibfnamefont {P.~A.}\ \bibnamefont
  {Butler}}\ and\ \bibinfo {author} {\bibfnamefont {W.}~\bibnamefont
  {Nazarewicz}},\ }\href {\doibase 10.1103/RevModPhys.68.349} {\bibfield
  {journal} {\bibinfo  {journal} {Rev. Mod. Phys.}\ }\textbf {\bibinfo {volume}
  {68}},\ \bibinfo {pages} {349} (\bibinfo {year} {1996})}\BibitemShut
  {NoStop}%
\bibitem [{\citenamefont {Butler}(2016)}]{Butler_2016}%
  \BibitemOpen
  \bibfield  {author} {\bibinfo {author} {\bibfnamefont {P.~A.}\ \bibnamefont
  {Butler}},\ }\href {\doibase 10.1088/0954-3899/43/7/073002} {\bibfield
  {journal} {\bibinfo  {journal} {J. Phys. G}\ }\textbf {\bibinfo {volume}
  {43}},\ \bibinfo {pages} {073002} (\bibinfo {year} {2016})}\BibitemShut
  {NoStop}%
\bibitem [{\citenamefont {Frauendorf}(2001)}]{Frauendorf2001Spontaneous}%
  \BibitemOpen
  \bibfield  {author} {\bibinfo {author} {\bibfnamefont {S.}~\bibnamefont
  {Frauendorf}},\ }\href {\doibase 10.1103/RevModPhys.73.463} {\bibfield
  {journal} {\bibinfo  {journal} {Rev. Mod. Phys.}\ }\textbf {\bibinfo {volume}
  {73}},\ \bibinfo {pages} {463} (\bibinfo {year} {2001})}\BibitemShut
  {NoStop}%
\bibitem [{\citenamefont {Meng}\ and\ \citenamefont
  {Zhang}(2010)}]{Meng2010Open}%
  \BibitemOpen
  \bibfield  {author} {\bibinfo {author} {\bibfnamefont {J.}~\bibnamefont
  {Meng}}\ and\ \bibinfo {author} {\bibfnamefont {S.~Q.}\ \bibnamefont
  {Zhang}},\ }\href {\doibase 10.1088/0954-3899/37/6/064025} {\bibfield
  {journal} {\bibinfo  {journal} {J. Phys. G}\ }\textbf {\bibinfo {volume}
  {37}},\ \bibinfo {pages} {064025} (\bibinfo {year} {2010})}\BibitemShut
  {NoStop}%
\bibitem [{\citenamefont {Meng}\ and\ \citenamefont
  {Zhao}(2016)}]{Meng2016Nuclear}%
  \BibitemOpen
  \bibfield  {author} {\bibinfo {author} {\bibfnamefont {J.}~\bibnamefont
  {Meng}}\ and\ \bibinfo {author} {\bibfnamefont {P.~W.}\ \bibnamefont
  {Zhao}},\ }\href {\doibase 10.1088/0031-8949/91/5/053008} {\bibfield
  {journal} {\bibinfo  {journal} {Phys. Scr.}\ }\textbf {\bibinfo {volume}
  {91}},\ \bibinfo {pages} {053008} (\bibinfo {year} {2016})}\BibitemShut
  {NoStop}%
\bibitem [{\citenamefont {Raduta}(2016)}]{Raduta2016Prog.Part.Nucl.Phys.241}%
  \BibitemOpen
  \bibfield  {author} {\bibinfo {author} {\bibfnamefont {A.~A.}\ \bibnamefont
  {Raduta}},\ }\href {\doibase 10.1016/j.ppnp.2016.05.002} {\bibfield
  {journal} {\bibinfo  {journal} {Prog. Part. Nucl. Phys.}\ }\textbf {\bibinfo
  {volume} {90}},\ \bibinfo {pages} {241} (\bibinfo {year} {2016})}\BibitemShut
  {NoStop}%
\bibitem [{\citenamefont {Frauendorf}(2018)}]{Frauendorf2018Phys.Scripta43003}%
  \BibitemOpen
  \bibfield  {author} {\bibinfo {author} {\bibfnamefont {S.}~\bibnamefont
  {Frauendorf}},\ }\href {\doibase 10.1088/1402-4896/aaa2e9} {\bibfield
  {journal} {\bibinfo  {journal} {Phys. Scr.}\ }\textbf {\bibinfo {volume}
  {93}},\ \bibinfo {pages} {043003} (\bibinfo {year} {2018})}\BibitemShut
  {NoStop}%
\bibitem [{\citenamefont {Xiong}\ and\ \citenamefont
  {Wang}(2019)}]{Xiong2019Atom.DataNucl.DataTabl.193}%
  \BibitemOpen
  \bibfield  {author} {\bibinfo {author} {\bibfnamefont {B.~W.}\ \bibnamefont
  {Xiong}}\ and\ \bibinfo {author} {\bibfnamefont {Y.~Y.}\ \bibnamefont
  {Wang}},\ }\href {\doibase 10.1016/j.adt.2018.05.002} {\bibfield  {journal}
  {\bibinfo  {journal} {Atom. Data Nucl. Data Tabl.}\ }\textbf {\bibinfo
  {volume} {125}},\ \bibinfo {pages} {193} (\bibinfo {year}
  {2019})}\BibitemShut {NoStop}%
\bibitem [{\citenamefont {Gaffney}\ \emph {et~al.}(2013)\citenamefont
  {Gaffney}, \citenamefont {Butler}, \citenamefont {Scheck}, \citenamefont
  {Hayes}, \citenamefont {Wenander}, \citenamefont {Albers}, \citenamefont
  {Bastin}, \citenamefont {Bauer}, \citenamefont {Blazhev}, \citenamefont
  {B{\"o}nig}, \citenamefont {Bree}, \citenamefont {Cederk{\"a}ll},
  \citenamefont {Chupp}, \citenamefont {Cline}, \citenamefont {Cocolios},
  \citenamefont {Davinson}, \citenamefont {De~Witte}, \citenamefont {Diriken},
  \citenamefont {Grahn}, \citenamefont {Herzan}, \citenamefont {Huyse},
  \citenamefont {Jenkins}, \citenamefont {Joss}, \citenamefont {Kesteloot},
  \citenamefont {Konkj}, \citenamefont {Kowalczyk}, \citenamefont {Kr{\"o}ll},
  \citenamefont {Kwan}, \citenamefont {Lutter}, \citenamefont {Moschner},
  \citenamefont {Napiorkowski}, \citenamefont {Pakarinen}, \citenamefont
  {Pfeiffer}, \citenamefont {Radeck}, \citenamefont {Reiter}, \citenamefont
  {Reynders}, \citenamefont {Rigby}, \citenamefont {Robledo}, \citenamefont
  {Rudigier}, \citenamefont {Sambi}, \citenamefont {Seidlitz}, \citenamefont
  {Siebeck}, \citenamefont {Stora}, \citenamefont {Thoele}, \citenamefont
  {Van~Duppen}, \citenamefont {Vermeulen}, \citenamefont {von Schmid},
  \citenamefont {Voulot}, \citenamefont {Warr}, \citenamefont {Wimmer},
  \citenamefont {Wrzosek-Lipska}, \citenamefont {Wu},\ and\ \citenamefont
  {Zielinska}}]{gaffney2013studies}%
  \BibitemOpen
  \bibfield  {author} {\bibinfo {author} {\bibfnamefont {L.~P.}\ \bibnamefont
  {Gaffney}}, \bibinfo {author} {\bibfnamefont {P.~A.}\ \bibnamefont {Butler}},
  \bibinfo {author} {\bibfnamefont {M.}~\bibnamefont {Scheck}}, \bibinfo
  {author} {\bibfnamefont {A.~B.}\ \bibnamefont {Hayes}}, \bibinfo {author}
  {\bibfnamefont {F.}~\bibnamefont {Wenander}}, \bibinfo {author}
  {\bibfnamefont {M.}~\bibnamefont {Albers}}, \bibinfo {author} {\bibfnamefont
  {B.}~\bibnamefont {Bastin}}, \bibinfo {author} {\bibfnamefont
  {C.}~\bibnamefont {Bauer}}, \bibinfo {author} {\bibfnamefont
  {A.}~\bibnamefont {Blazhev}}, \bibinfo {author} {\bibfnamefont
  {S.}~\bibnamefont {B{\"o}nig}}, \bibinfo {author} {\bibfnamefont
  {N.}~\bibnamefont {Bree}}, \bibinfo {author} {\bibfnamefont {J.}~\bibnamefont
  {Cederk{\"a}ll}}, \bibinfo {author} {\bibfnamefont {J.}~\bibnamefont
  {Chupp}}, \bibinfo {author} {\bibfnamefont {D.}~\bibnamefont {Cline}},
  \bibinfo {author} {\bibfnamefont {T.~E.}\ \bibnamefont {Cocolios}}, \bibinfo
  {author} {\bibfnamefont {T.}~\bibnamefont {Davinson}}, \bibinfo {author}
  {\bibfnamefont {H.}~\bibnamefont {De~Witte}}, \bibinfo {author}
  {\bibfnamefont {J.}~\bibnamefont {Diriken}}, \bibinfo {author} {\bibfnamefont
  {T.}~\bibnamefont {Grahn}}, \bibinfo {author} {\bibfnamefont
  {A.}~\bibnamefont {Herzan}}, \bibinfo {author} {\bibfnamefont
  {M.}~\bibnamefont {Huyse}}, \bibinfo {author} {\bibfnamefont {D.~G.}\
  \bibnamefont {Jenkins}}, \bibinfo {author} {\bibfnamefont {D.~T.}\
  \bibnamefont {Joss}}, \bibinfo {author} {\bibfnamefont {N.}~\bibnamefont
  {Kesteloot}}, \bibinfo {author} {\bibfnamefont {J.}~\bibnamefont {Konkj}},
  \bibinfo {author} {\bibfnamefont {M.}~\bibnamefont {Kowalczyk}}, \bibinfo
  {author} {\bibfnamefont {T.}~\bibnamefont {Kr{\"o}ll}}, \bibinfo {author}
  {\bibfnamefont {E.}~\bibnamefont {Kwan}}, \bibinfo {author} {\bibfnamefont
  {R.}~\bibnamefont {Lutter}}, \bibinfo {author} {\bibfnamefont
  {K.}~\bibnamefont {Moschner}}, \bibinfo {author} {\bibfnamefont
  {P.}~\bibnamefont {Napiorkowski}}, \bibinfo {author} {\bibfnamefont
  {J.}~\bibnamefont {Pakarinen}}, \bibinfo {author} {\bibfnamefont
  {M.}~\bibnamefont {Pfeiffer}}, \bibinfo {author} {\bibfnamefont
  {D.}~\bibnamefont {Radeck}}, \bibinfo {author} {\bibfnamefont
  {P.}~\bibnamefont {Reiter}}, \bibinfo {author} {\bibfnamefont
  {K.}~\bibnamefont {Reynders}}, \bibinfo {author} {\bibfnamefont {S.~V.}\
  \bibnamefont {Rigby}}, \bibinfo {author} {\bibfnamefont {L.~M.}\ \bibnamefont
  {Robledo}}, \bibinfo {author} {\bibfnamefont {M.}~\bibnamefont {Rudigier}},
  \bibinfo {author} {\bibfnamefont {S.}~\bibnamefont {Sambi}}, \bibinfo
  {author} {\bibfnamefont {M.}~\bibnamefont {Seidlitz}}, \bibinfo {author}
  {\bibfnamefont {B.}~\bibnamefont {Siebeck}}, \bibinfo {author} {\bibfnamefont
  {T.}~\bibnamefont {Stora}}, \bibinfo {author} {\bibfnamefont
  {P.}~\bibnamefont {Thoele}}, \bibinfo {author} {\bibfnamefont
  {P.}~\bibnamefont {Van~Duppen}}, \bibinfo {author} {\bibfnamefont {M.~J.}\
  \bibnamefont {Vermeulen}}, \bibinfo {author} {\bibfnamefont {M.}~\bibnamefont
  {von Schmid}}, \bibinfo {author} {\bibfnamefont {D.}~\bibnamefont {Voulot}},
  \bibinfo {author} {\bibfnamefont {N.}~\bibnamefont {Warr}}, \bibinfo {author}
  {\bibfnamefont {K.}~\bibnamefont {Wimmer}}, \bibinfo {author} {\bibfnamefont
  {K.}~\bibnamefont {Wrzosek-Lipska}}, \bibinfo {author} {\bibfnamefont
  {C.~Y.}\ \bibnamefont {Wu}}, \ and\ \bibinfo {author} {\bibfnamefont
  {M.}~\bibnamefont {Zielinska}},\ }\href {\doibase 10.1038/nature12073}
  {\bibfield  {journal} {\bibinfo  {journal} {Nature}\ }\textbf {\bibinfo
  {volume} {497}},\ \bibinfo {pages} {199} (\bibinfo {year}
  {2013})}\BibitemShut {NoStop}%
\bibitem [{\citenamefont {Meng}\ \emph {et~al.}(2006)\citenamefont {Meng},
  \citenamefont {Peng}, \citenamefont {Zhang},\ and\ \citenamefont
  {Zhou}}]{Meng2006Possible}%
  \BibitemOpen
  \bibfield  {author} {\bibinfo {author} {\bibfnamefont {J.}~\bibnamefont
  {Meng}}, \bibinfo {author} {\bibfnamefont {J.}~\bibnamefont {Peng}}, \bibinfo
  {author} {\bibfnamefont {S.~Q.}\ \bibnamefont {Zhang}}, \ and\ \bibinfo
  {author} {\bibfnamefont {S.-G.}\ \bibnamefont {Zhou}},\ }\href {\doibase
  10.1103/PhysRevC.73.037303} {\bibfield  {journal} {\bibinfo  {journal} {Phys.
  Rev. C}\ }\textbf {\bibinfo {volume} {73}},\ \bibinfo {pages} {037303}
  (\bibinfo {year} {2006})}\BibitemShut {NoStop}%
\bibitem [{\citenamefont {Zhao}(2017)}]{Zhao2017Phys.Lett.B1}%
  \BibitemOpen
  \bibfield  {author} {\bibinfo {author} {\bibfnamefont {P.~W.}\ \bibnamefont
  {Zhao}},\ }\href {\doibase 10.1016/j.physletb.2017.08.001} {\bibfield
  {journal} {\bibinfo  {journal} {Phys. Lett. B}\ }\textbf {\bibinfo {volume}
  {773}},\ \bibinfo {pages} {1} (\bibinfo {year} {2017})}\BibitemShut {NoStop}%
\bibitem [{\citenamefont {Ayangeakaa}\ \emph {et~al.}(2013)\citenamefont
  {Ayangeakaa}, \citenamefont {Garg}, \citenamefont {Anthony}, \citenamefont
  {Frauendorf}, \citenamefont {Matta}, \citenamefont {Nayak}, \citenamefont
  {Patel}, \citenamefont {Chen}, \citenamefont {Zhang}, \citenamefont {Zhao},
  \citenamefont {Qi}, \citenamefont {Meng}, \citenamefont {Janssens},
  \citenamefont {Carpenter}, \citenamefont {Chiara}, \citenamefont {Kondev},
  \citenamefont {Lauritsen}, \citenamefont {Seweryniak}, \citenamefont {Zhu},
  \citenamefont {Ghugre},\ and\ \citenamefont
  {Palit}}]{Ayangeakaa2013Evidence}%
  \BibitemOpen
  \bibfield  {author} {\bibinfo {author} {\bibfnamefont {A.~D.}\ \bibnamefont
  {Ayangeakaa}}, \bibinfo {author} {\bibfnamefont {U.}~\bibnamefont {Garg}},
  \bibinfo {author} {\bibfnamefont {M.~D.}\ \bibnamefont {Anthony}}, \bibinfo
  {author} {\bibfnamefont {S.}~\bibnamefont {Frauendorf}}, \bibinfo {author}
  {\bibfnamefont {J.~T.}\ \bibnamefont {Matta}}, \bibinfo {author}
  {\bibfnamefont {B.~K.}\ \bibnamefont {Nayak}}, \bibinfo {author}
  {\bibfnamefont {D.}~\bibnamefont {Patel}}, \bibinfo {author} {\bibfnamefont
  {Q.~B.}\ \bibnamefont {Chen}}, \bibinfo {author} {\bibfnamefont {S.~Q.}\
  \bibnamefont {Zhang}}, \bibinfo {author} {\bibfnamefont {P.~W.}\ \bibnamefont
  {Zhao}}, \bibinfo {author} {\bibfnamefont {B.}~\bibnamefont {Qi}}, \bibinfo
  {author} {\bibfnamefont {J.}~\bibnamefont {Meng}}, \bibinfo {author}
  {\bibfnamefont {R.~V.~F.}\ \bibnamefont {Janssens}}, \bibinfo {author}
  {\bibfnamefont {M.~P.}\ \bibnamefont {Carpenter}}, \bibinfo {author}
  {\bibfnamefont {C.~J.}\ \bibnamefont {Chiara}}, \bibinfo {author}
  {\bibfnamefont {F.~G.}\ \bibnamefont {Kondev}}, \bibinfo {author}
  {\bibfnamefont {T.}~\bibnamefont {Lauritsen}}, \bibinfo {author}
  {\bibfnamefont {D.}~\bibnamefont {Seweryniak}}, \bibinfo {author}
  {\bibfnamefont {S.}~\bibnamefont {Zhu}}, \bibinfo {author} {\bibfnamefont
  {S.~S.}\ \bibnamefont {Ghugre}}, \ and\ \bibinfo {author} {\bibfnamefont
  {R.}~\bibnamefont {Palit}},\ }\href {\doibase 10.1103/PhysRevLett.110.172504}
  {\bibfield  {journal} {\bibinfo  {journal} {Phys. Rev. Lett.}\ }\textbf
  {\bibinfo {volume} {110}},\ \bibinfo {pages} {172504} (\bibinfo {year}
  {2013})}\BibitemShut {NoStop}%
\bibitem [{\citenamefont {Kuti}\ \emph {et~al.}(2014)\citenamefont {Kuti},
  \citenamefont {Chen}, \citenamefont {Tim\'ar}, \citenamefont {Sohler},
  \citenamefont {Zhang}, \citenamefont {Zhang}, \citenamefont {Zhao},
  \citenamefont {Meng}, \citenamefont {Starosta}, \citenamefont {Koike},
  \citenamefont {Paul}, \citenamefont {Fossan},\ and\ \citenamefont
  {Vaman}}]{Kuti2014Multiple}%
  \BibitemOpen
  \bibfield  {author} {\bibinfo {author} {\bibfnamefont {I.}~\bibnamefont
  {Kuti}}, \bibinfo {author} {\bibfnamefont {Q.~B.}\ \bibnamefont {Chen}},
  \bibinfo {author} {\bibfnamefont {J.}~\bibnamefont {Tim\'ar}}, \bibinfo
  {author} {\bibfnamefont {D.}~\bibnamefont {Sohler}}, \bibinfo {author}
  {\bibfnamefont {S.~Q.}\ \bibnamefont {Zhang}}, \bibinfo {author}
  {\bibfnamefont {Z.~H.}\ \bibnamefont {Zhang}}, \bibinfo {author}
  {\bibfnamefont {P.~W.}\ \bibnamefont {Zhao}}, \bibinfo {author}
  {\bibfnamefont {J.}~\bibnamefont {Meng}}, \bibinfo {author} {\bibfnamefont
  {K.}~\bibnamefont {Starosta}}, \bibinfo {author} {\bibfnamefont
  {T.}~\bibnamefont {Koike}}, \bibinfo {author} {\bibfnamefont {E.~S.}\
  \bibnamefont {Paul}}, \bibinfo {author} {\bibfnamefont {D.~B.}\ \bibnamefont
  {Fossan}}, \ and\ \bibinfo {author} {\bibfnamefont {C.}~\bibnamefont
  {Vaman}},\ }\href {\doibase 10.1103/PhysRevLett.113.032501} {\bibfield
  {journal} {\bibinfo  {journal} {Phys. Rev. Lett.}\ }\textbf {\bibinfo
  {volume} {113}},\ \bibinfo {pages} {032501} (\bibinfo {year}
  {2014})}\BibitemShut {NoStop}%
\bibitem [{\citenamefont {Liu}\ \emph {et~al.}(2016)\citenamefont {Liu},
  \citenamefont {Wang}, \citenamefont {Bark}, \citenamefont {Zhang},
  \citenamefont {Meng}, \citenamefont {Qi}, \citenamefont {Jones},
  \citenamefont {Wyngaardt}, \citenamefont {Zhao}, \citenamefont {Xu},
  \citenamefont {Zhou}, \citenamefont {Wang}, \citenamefont {Sun},
  \citenamefont {Liu}, \citenamefont {Li}, \citenamefont {Zhang}, \citenamefont
  {Jia}, \citenamefont {Li}, \citenamefont {Hua}, \citenamefont {Chen},
  \citenamefont {Xiao}, \citenamefont {Li}, \citenamefont {Zhu}, \citenamefont
  {Bucher}, \citenamefont {Dinoko}, \citenamefont {Easton}, \citenamefont
  {Juh\'asz}, \citenamefont {Kamblawe}, \citenamefont {Khaleel}, \citenamefont
  {Khumalo}, \citenamefont {Lawrie}, \citenamefont {Lawrie}, \citenamefont
  {Majola}, \citenamefont {Mullins}, \citenamefont {Murray}, \citenamefont
  {Ndayishimye}, \citenamefont {Negi}, \citenamefont {Noncolela}, \citenamefont
  {Ntshangase}, \citenamefont {Nyak\'o}, \citenamefont {Orce}, \citenamefont
  {Papka}, \citenamefont {Sharpey-Schafer}, \citenamefont {Shirinda},
  \citenamefont {Sithole}, \citenamefont {Stankiewicz},\ and\ \citenamefont
  {Wiedeking}}]{Liu2016Phys.Rev.Lett.112501}%
  \BibitemOpen
  \bibfield  {author} {\bibinfo {author} {\bibfnamefont {C.}~\bibnamefont
  {Liu}}, \bibinfo {author} {\bibfnamefont {S.~Y.}\ \bibnamefont {Wang}},
  \bibinfo {author} {\bibfnamefont {R.~A.}\ \bibnamefont {Bark}}, \bibinfo
  {author} {\bibfnamefont {S.~Q.}\ \bibnamefont {Zhang}}, \bibinfo {author}
  {\bibfnamefont {J.}~\bibnamefont {Meng}}, \bibinfo {author} {\bibfnamefont
  {B.}~\bibnamefont {Qi}}, \bibinfo {author} {\bibfnamefont {P.}~\bibnamefont
  {Jones}}, \bibinfo {author} {\bibfnamefont {S.~M.}\ \bibnamefont
  {Wyngaardt}}, \bibinfo {author} {\bibfnamefont {J.}~\bibnamefont {Zhao}},
  \bibinfo {author} {\bibfnamefont {C.}~\bibnamefont {Xu}}, \bibinfo {author}
  {\bibfnamefont {S.-G.}\ \bibnamefont {Zhou}}, \bibinfo {author}
  {\bibfnamefont {S.}~\bibnamefont {Wang}}, \bibinfo {author} {\bibfnamefont
  {D.~P.}\ \bibnamefont {Sun}}, \bibinfo {author} {\bibfnamefont
  {L.}~\bibnamefont {Liu}}, \bibinfo {author} {\bibfnamefont {Z.~Q.}\
  \bibnamefont {Li}}, \bibinfo {author} {\bibfnamefont {N.~B.}\ \bibnamefont
  {Zhang}}, \bibinfo {author} {\bibfnamefont {H.}~\bibnamefont {Jia}}, \bibinfo
  {author} {\bibfnamefont {X.~Q.}\ \bibnamefont {Li}}, \bibinfo {author}
  {\bibfnamefont {H.}~\bibnamefont {Hua}}, \bibinfo {author} {\bibfnamefont
  {Q.~B.}\ \bibnamefont {Chen}}, \bibinfo {author} {\bibfnamefont {Z.~G.}\
  \bibnamefont {Xiao}}, \bibinfo {author} {\bibfnamefont {H.~J.}\ \bibnamefont
  {Li}}, \bibinfo {author} {\bibfnamefont {L.~H.}\ \bibnamefont {Zhu}},
  \bibinfo {author} {\bibfnamefont {T.~D.}\ \bibnamefont {Bucher}}, \bibinfo
  {author} {\bibfnamefont {T.}~\bibnamefont {Dinoko}}, \bibinfo {author}
  {\bibfnamefont {J.}~\bibnamefont {Easton}}, \bibinfo {author} {\bibfnamefont
  {K.}~\bibnamefont {Juh\'asz}}, \bibinfo {author} {\bibfnamefont
  {A.}~\bibnamefont {Kamblawe}}, \bibinfo {author} {\bibfnamefont
  {E.}~\bibnamefont {Khaleel}}, \bibinfo {author} {\bibfnamefont
  {N.}~\bibnamefont {Khumalo}}, \bibinfo {author} {\bibfnamefont {E.~A.}\
  \bibnamefont {Lawrie}}, \bibinfo {author} {\bibfnamefont {J.~J.}\
  \bibnamefont {Lawrie}}, \bibinfo {author} {\bibfnamefont {S.~N.~T.}\
  \bibnamefont {Majola}}, \bibinfo {author} {\bibfnamefont {S.~M.}\
  \bibnamefont {Mullins}}, \bibinfo {author} {\bibfnamefont {S.}~\bibnamefont
  {Murray}}, \bibinfo {author} {\bibfnamefont {J.}~\bibnamefont {Ndayishimye}},
  \bibinfo {author} {\bibfnamefont {D.}~\bibnamefont {Negi}}, \bibinfo {author}
  {\bibfnamefont {S.~P.}\ \bibnamefont {Noncolela}}, \bibinfo {author}
  {\bibfnamefont {S.~S.}\ \bibnamefont {Ntshangase}}, \bibinfo {author}
  {\bibfnamefont {B.~M.}\ \bibnamefont {Nyak\'o}}, \bibinfo {author}
  {\bibfnamefont {J.~N.}\ \bibnamefont {Orce}}, \bibinfo {author}
  {\bibfnamefont {P.}~\bibnamefont {Papka}}, \bibinfo {author} {\bibfnamefont
  {J.~F.}\ \bibnamefont {Sharpey-Schafer}}, \bibinfo {author} {\bibfnamefont
  {O.}~\bibnamefont {Shirinda}}, \bibinfo {author} {\bibfnamefont
  {P.}~\bibnamefont {Sithole}}, \bibinfo {author} {\bibfnamefont {M.~A.}\
  \bibnamefont {Stankiewicz}}, \ and\ \bibinfo {author} {\bibfnamefont
  {M.}~\bibnamefont {Wiedeking}},\ }\href {\doibase
  10.1103/PhysRevLett.116.112501} {\bibfield  {journal} {\bibinfo  {journal}
  {Phys. Rev. Lett.}\ }\textbf {\bibinfo {volume} {116}},\ \bibinfo {pages}
  {112501} (\bibinfo {year} {2016})}\BibitemShut {NoStop}%
\bibitem [{\citenamefont {Petrache}\ \emph {et~al.}(2018)\citenamefont
  {Petrache}, \citenamefont {Lv}, \citenamefont {Astier}, \citenamefont
  {Dupont}, \citenamefont {Wang}, \citenamefont {Zhang}, \citenamefont {Zhao},
  \citenamefont {Ren}, \citenamefont {Meng}, \citenamefont {Greenlees},
  \citenamefont {Badran}, \citenamefont {Cox}, \citenamefont {Grahn},
  \citenamefont {Julin}, \citenamefont {Juutinen}, \citenamefont {Konki},
  \citenamefont {Pakarinen}, \citenamefont {Papadakis}, \citenamefont
  {Partanen}, \citenamefont {Rahkila}, \citenamefont {Sandzelius},
  \citenamefont {Saren}, \citenamefont {Scholey}, \citenamefont {Sorri},
  \citenamefont {Stolze}, \citenamefont {Uusitalo}, \citenamefont {Cederwall},
  \citenamefont {Aktas}, \citenamefont {Ertoprak}, \citenamefont {Liu},
  \citenamefont {Matta}, \citenamefont {Subramaniam}, \citenamefont {Guo},
  \citenamefont {Liu}, \citenamefont {Zhou}, \citenamefont {Wang},
  \citenamefont {Kuti}, \citenamefont {Tim\'ar}, \citenamefont {Tucholski},
  \citenamefont {Srebrny},\ and\ \citenamefont
  {Andreoiu}}]{Petrache2018Phys.Rev.C41304}%
  \BibitemOpen
  \bibfield  {author} {\bibinfo {author} {\bibfnamefont {C.~M.}\ \bibnamefont
  {Petrache}}, \bibinfo {author} {\bibfnamefont {B.~F.}\ \bibnamefont {Lv}},
  \bibinfo {author} {\bibfnamefont {A.}~\bibnamefont {Astier}}, \bibinfo
  {author} {\bibfnamefont {E.}~\bibnamefont {Dupont}}, \bibinfo {author}
  {\bibfnamefont {Y.~K.}\ \bibnamefont {Wang}}, \bibinfo {author}
  {\bibfnamefont {S.~Q.}\ \bibnamefont {Zhang}}, \bibinfo {author}
  {\bibfnamefont {P.~W.}\ \bibnamefont {Zhao}}, \bibinfo {author}
  {\bibfnamefont {Z.~X.}\ \bibnamefont {Ren}}, \bibinfo {author} {\bibfnamefont
  {J.}~\bibnamefont {Meng}}, \bibinfo {author} {\bibfnamefont {P.~T.}\
  \bibnamefont {Greenlees}}, \bibinfo {author} {\bibfnamefont {H.}~\bibnamefont
  {Badran}}, \bibinfo {author} {\bibfnamefont {D.~M.}\ \bibnamefont {Cox}},
  \bibinfo {author} {\bibfnamefont {T.}~\bibnamefont {Grahn}}, \bibinfo
  {author} {\bibfnamefont {R.}~\bibnamefont {Julin}}, \bibinfo {author}
  {\bibfnamefont {S.}~\bibnamefont {Juutinen}}, \bibinfo {author}
  {\bibfnamefont {J.}~\bibnamefont {Konki}}, \bibinfo {author} {\bibfnamefont
  {J.}~\bibnamefont {Pakarinen}}, \bibinfo {author} {\bibfnamefont
  {P.}~\bibnamefont {Papadakis}}, \bibinfo {author} {\bibfnamefont
  {J.}~\bibnamefont {Partanen}}, \bibinfo {author} {\bibfnamefont
  {P.}~\bibnamefont {Rahkila}}, \bibinfo {author} {\bibfnamefont
  {M.}~\bibnamefont {Sandzelius}}, \bibinfo {author} {\bibfnamefont
  {J.}~\bibnamefont {Saren}}, \bibinfo {author} {\bibfnamefont
  {C.}~\bibnamefont {Scholey}}, \bibinfo {author} {\bibfnamefont
  {J.}~\bibnamefont {Sorri}}, \bibinfo {author} {\bibfnamefont
  {S.}~\bibnamefont {Stolze}}, \bibinfo {author} {\bibfnamefont
  {J.}~\bibnamefont {Uusitalo}}, \bibinfo {author} {\bibfnamefont
  {B.}~\bibnamefont {Cederwall}}, \bibinfo {author} {\bibfnamefont
  {{\"O}.}~\bibnamefont {Aktas}}, \bibinfo {author} {\bibfnamefont
  {A.}~\bibnamefont {Ertoprak}}, \bibinfo {author} {\bibfnamefont
  {H.}~\bibnamefont {Liu}}, \bibinfo {author} {\bibfnamefont {S.}~\bibnamefont
  {Matta}}, \bibinfo {author} {\bibfnamefont {P.}~\bibnamefont {Subramaniam}},
  \bibinfo {author} {\bibfnamefont {S.}~\bibnamefont {Guo}}, \bibinfo {author}
  {\bibfnamefont {M.~L.}\ \bibnamefont {Liu}}, \bibinfo {author} {\bibfnamefont
  {X.~H.}\ \bibnamefont {Zhou}}, \bibinfo {author} {\bibfnamefont {K.~L.}\
  \bibnamefont {Wang}}, \bibinfo {author} {\bibfnamefont {I.}~\bibnamefont
  {Kuti}}, \bibinfo {author} {\bibfnamefont {J.}~\bibnamefont {Tim\'ar}},
  \bibinfo {author} {\bibfnamefont {A.}~\bibnamefont {Tucholski}}, \bibinfo
  {author} {\bibfnamefont {J.}~\bibnamefont {Srebrny}}, \ and\ \bibinfo
  {author} {\bibfnamefont {C.}~\bibnamefont {Andreoiu}},\ }\href {\doibase
  10.1103/PhysRevC.97.041304} {\bibfield  {journal} {\bibinfo  {journal} {Phys.
  Rev. C}\ }\textbf {\bibinfo {volume} {97}},\ \bibinfo {pages} {041304(R)}
  (\bibinfo {year} {2018})}\BibitemShut {NoStop}%
\bibitem [{\citenamefont {Roy}\ \emph {et~al.}(2018)\citenamefont {Roy},
  \citenamefont {Mukherjee}, \citenamefont {Asgar}, \citenamefont
  {Bhattacharyya}, \citenamefont {Bhattacharya}, \citenamefont {Bhattacharya},
  \citenamefont {Bhattacharya}, \citenamefont {Ghosh}, \citenamefont
  {Banerjee}, \citenamefont {Kundu}, \citenamefont {Rana}, \citenamefont {Roy},
  \citenamefont {Pandey}, \citenamefont {Meena}, \citenamefont {Dhal},
  \citenamefont {Palit}, \citenamefont {Saha}, \citenamefont {Sethi},
  \citenamefont {Thakur}, \citenamefont {Naidu}, \citenamefont {Jadav},
  \citenamefont {Dhonti}, \citenamefont {Pai},\ and\ \citenamefont
  {Goswami}}]{Roy2018Phys.Lett.768}%
  \BibitemOpen
  \bibfield  {author} {\bibinfo {author} {\bibfnamefont {T.}~\bibnamefont
  {Roy}}, \bibinfo {author} {\bibfnamefont {G.}~\bibnamefont {Mukherjee}},
  \bibinfo {author} {\bibfnamefont {M.}~\bibnamefont {Asgar}}, \bibinfo
  {author} {\bibfnamefont {S.}~\bibnamefont {Bhattacharyya}}, \bibinfo {author}
  {\bibfnamefont {S.}~\bibnamefont {Bhattacharya}}, \bibinfo {author}
  {\bibfnamefont {C.}~\bibnamefont {Bhattacharya}}, \bibinfo {author}
  {\bibfnamefont {S.}~\bibnamefont {Bhattacharya}}, \bibinfo {author}
  {\bibfnamefont {T.}~\bibnamefont {Ghosh}}, \bibinfo {author} {\bibfnamefont
  {K.}~\bibnamefont {Banerjee}}, \bibinfo {author} {\bibfnamefont
  {S.}~\bibnamefont {Kundu}}, \bibinfo {author} {\bibfnamefont
  {T.}~\bibnamefont {Rana}}, \bibinfo {author} {\bibfnamefont {P.}~\bibnamefont
  {Roy}}, \bibinfo {author} {\bibfnamefont {R.}~\bibnamefont {Pandey}},
  \bibinfo {author} {\bibfnamefont {J.}~\bibnamefont {Meena}}, \bibinfo
  {author} {\bibfnamefont {A.}~\bibnamefont {Dhal}}, \bibinfo {author}
  {\bibfnamefont {R.}~\bibnamefont {Palit}}, \bibinfo {author} {\bibfnamefont
  {S.}~\bibnamefont {Saha}}, \bibinfo {author} {\bibfnamefont {J.}~\bibnamefont
  {Sethi}}, \bibinfo {author} {\bibfnamefont {S.}~\bibnamefont {Thakur}},
  \bibinfo {author} {\bibfnamefont {B.}~\bibnamefont {Naidu}}, \bibinfo
  {author} {\bibfnamefont {S.}~\bibnamefont {Jadav}}, \bibinfo {author}
  {\bibfnamefont {R.}~\bibnamefont {Dhonti}}, \bibinfo {author} {\bibfnamefont
  {H.}~\bibnamefont {Pai}}, \ and\ \bibinfo {author} {\bibfnamefont
  {A.}~\bibnamefont {Goswami}},\ }\href {\doibase
  10.1016/j.physletb.2018.06.033} {\bibfield  {journal} {\bibinfo  {journal}
  {Phys. Lett. B}\ }\textbf {\bibinfo {volume} {782}},\ \bibinfo {pages} {768}
  (\bibinfo {year} {2018})}\BibitemShut {NoStop}%
\bibitem [{\citenamefont {Zhu}\ \emph {et~al.}(2003)\citenamefont {Zhu},
  \citenamefont {Garg}, \citenamefont {Nayak}, \citenamefont {Ghugre},
  \citenamefont {Pattabiraman}, \citenamefont {Fossan}, \citenamefont {Koike},
  \citenamefont {Starosta}, \citenamefont {Vaman}, \citenamefont {Janssens},
  \citenamefont {Chakrawarthy}, \citenamefont {Whitehead}, \citenamefont
  {Macchiavelli},\ and\ \citenamefont {Frauendorf}}]{Zhu2003A}%
  \BibitemOpen
  \bibfield  {author} {\bibinfo {author} {\bibfnamefont {S.}~\bibnamefont
  {Zhu}}, \bibinfo {author} {\bibfnamefont {U.}~\bibnamefont {Garg}}, \bibinfo
  {author} {\bibfnamefont {B.~K.}\ \bibnamefont {Nayak}}, \bibinfo {author}
  {\bibfnamefont {S.~S.}\ \bibnamefont {Ghugre}}, \bibinfo {author}
  {\bibfnamefont {N.~S.}\ \bibnamefont {Pattabiraman}}, \bibinfo {author}
  {\bibfnamefont {D.~B.}\ \bibnamefont {Fossan}}, \bibinfo {author}
  {\bibfnamefont {T.}~\bibnamefont {Koike}}, \bibinfo {author} {\bibfnamefont
  {K.}~\bibnamefont {Starosta}}, \bibinfo {author} {\bibfnamefont
  {C.}~\bibnamefont {Vaman}}, \bibinfo {author} {\bibfnamefont {R.~V.~F.}\
  \bibnamefont {Janssens}}, \bibinfo {author} {\bibfnamefont {R.~S.}\
  \bibnamefont {Chakrawarthy}}, \bibinfo {author} {\bibfnamefont
  {M.}~\bibnamefont {Whitehead}}, \bibinfo {author} {\bibfnamefont {A.~O.}\
  \bibnamefont {Macchiavelli}}, \ and\ \bibinfo {author} {\bibfnamefont
  {S.}~\bibnamefont {Frauendorf}},\ }\href {\doibase
  10.1103/PhysRevLett.91.132501} {\bibfield  {journal} {\bibinfo  {journal}
  {Phys. Rev. Lett.}\ }\textbf {\bibinfo {volume} {91}},\ \bibinfo {pages}
  {132501} (\bibinfo {year} {2003})}\BibitemShut {NoStop}%
\bibitem [{\citenamefont {Mukhopadhyay}\ \emph {et~al.}(2007)\citenamefont
  {Mukhopadhyay}, \citenamefont {Almehed}, \citenamefont {Garg}, \citenamefont
  {Frauendorf}, \citenamefont {Li}, \citenamefont {Madhusudhana~Rao},
  \citenamefont {Wang}, \citenamefont {Ghugre}, \citenamefont {Carpenter},
  \citenamefont {Gros}, \citenamefont {Hecht}, \citenamefont {Janssens},
  \citenamefont {Kondev}, \citenamefont {Lauritsen}, \citenamefont
  {Seweryniak},\ and\ \citenamefont {Zhu}}]{Mukhopadhyay2007From}%
  \BibitemOpen
  \bibfield  {author} {\bibinfo {author} {\bibfnamefont {S.}~\bibnamefont
  {Mukhopadhyay}}, \bibinfo {author} {\bibfnamefont {D.}~\bibnamefont
  {Almehed}}, \bibinfo {author} {\bibfnamefont {U.}~\bibnamefont {Garg}},
  \bibinfo {author} {\bibfnamefont {S.}~\bibnamefont {Frauendorf}}, \bibinfo
  {author} {\bibfnamefont {T.}~\bibnamefont {Li}}, \bibinfo {author}
  {\bibfnamefont {P.~V.}\ \bibnamefont {Madhusudhana~Rao}}, \bibinfo {author}
  {\bibfnamefont {X.}~\bibnamefont {Wang}}, \bibinfo {author} {\bibfnamefont
  {S.~S.}\ \bibnamefont {Ghugre}}, \bibinfo {author} {\bibfnamefont {M.~P.}\
  \bibnamefont {Carpenter}}, \bibinfo {author} {\bibfnamefont {S.}~\bibnamefont
  {Gros}}, \bibinfo {author} {\bibfnamefont {A.}~\bibnamefont {Hecht}},
  \bibinfo {author} {\bibfnamefont {R.~V.~F.}\ \bibnamefont {Janssens}},
  \bibinfo {author} {\bibfnamefont {F.~G.}\ \bibnamefont {Kondev}}, \bibinfo
  {author} {\bibfnamefont {T.}~\bibnamefont {Lauritsen}}, \bibinfo {author}
  {\bibfnamefont {D.}~\bibnamefont {Seweryniak}}, \ and\ \bibinfo {author}
  {\bibfnamefont {S.}~\bibnamefont {Zhu}},\ }\href {\doibase
  10.1103/PhysRevLett.99.172501} {\bibfield  {journal} {\bibinfo  {journal}
  {Phys. Rev. Lett.}\ }\textbf {\bibinfo {volume} {99}},\ \bibinfo {pages}
  {172501} (\bibinfo {year} {2007})}\BibitemShut {NoStop}%
\bibitem [{\citenamefont {Lv}\ \emph {et~al.}(2019)\citenamefont {Lv},
  \citenamefont {Petrache}, \citenamefont {Chen}, \citenamefont {Meng},
  \citenamefont {Astier}, \citenamefont {Dupont}, \citenamefont {Greenlees},
  \citenamefont {Badran}, \citenamefont {Calverley}, \citenamefont {Cox},
  \citenamefont {Grahn}, \citenamefont {Hilton}, \citenamefont {Julin},
  \citenamefont {Juutinen}, \citenamefont {Konki}, \citenamefont {Pakarinen},
  \citenamefont {Papadakis}, \citenamefont {Partanen}, \citenamefont {Rahkila},
  \citenamefont {Ruotsalainen}, \citenamefont {Sandzelius}, \citenamefont
  {Saren}, \citenamefont {Scholey}, \citenamefont {Sorri}, \citenamefont
  {Stolze}, \citenamefont {Uusitalo}, \citenamefont {Cederwall}, \citenamefont
  {Ertoprak}, \citenamefont {Liu}, \citenamefont {Guo}, \citenamefont {Liu},
  \citenamefont {Wang}, \citenamefont {Zhou}, \citenamefont {Kuti},
  \citenamefont {Tim\'ar}, \citenamefont {Tucholski}, \citenamefont {Srebrny},\
  and\ \citenamefont {Andreoiu}}]{PhysRevC.100.024314}%
  \BibitemOpen
  \bibfield  {author} {\bibinfo {author} {\bibfnamefont {B.~F.}\ \bibnamefont
  {Lv}}, \bibinfo {author} {\bibfnamefont {C.~M.}\ \bibnamefont {Petrache}},
  \bibinfo {author} {\bibfnamefont {Q.~B.}\ \bibnamefont {Chen}}, \bibinfo
  {author} {\bibfnamefont {J.}~\bibnamefont {Meng}}, \bibinfo {author}
  {\bibfnamefont {A.}~\bibnamefont {Astier}}, \bibinfo {author} {\bibfnamefont
  {E.}~\bibnamefont {Dupont}}, \bibinfo {author} {\bibfnamefont
  {P.}~\bibnamefont {Greenlees}}, \bibinfo {author} {\bibfnamefont
  {H.}~\bibnamefont {Badran}}, \bibinfo {author} {\bibfnamefont
  {T.}~\bibnamefont {Calverley}}, \bibinfo {author} {\bibfnamefont {D.~M.}\
  \bibnamefont {Cox}}, \bibinfo {author} {\bibfnamefont {T.}~\bibnamefont
  {Grahn}}, \bibinfo {author} {\bibfnamefont {J.}~\bibnamefont {Hilton}},
  \bibinfo {author} {\bibfnamefont {R.}~\bibnamefont {Julin}}, \bibinfo
  {author} {\bibfnamefont {S.}~\bibnamefont {Juutinen}}, \bibinfo {author}
  {\bibfnamefont {J.}~\bibnamefont {Konki}}, \bibinfo {author} {\bibfnamefont
  {J.}~\bibnamefont {Pakarinen}}, \bibinfo {author} {\bibfnamefont
  {P.}~\bibnamefont {Papadakis}}, \bibinfo {author} {\bibfnamefont
  {J.}~\bibnamefont {Partanen}}, \bibinfo {author} {\bibfnamefont
  {P.}~\bibnamefont {Rahkila}}, \bibinfo {author} {\bibfnamefont
  {P.}~\bibnamefont {Ruotsalainen}}, \bibinfo {author} {\bibfnamefont
  {M.}~\bibnamefont {Sandzelius}}, \bibinfo {author} {\bibfnamefont
  {J.}~\bibnamefont {Saren}}, \bibinfo {author} {\bibfnamefont
  {C.}~\bibnamefont {Scholey}}, \bibinfo {author} {\bibfnamefont
  {J.}~\bibnamefont {Sorri}}, \bibinfo {author} {\bibfnamefont
  {S.}~\bibnamefont {Stolze}}, \bibinfo {author} {\bibfnamefont
  {J.}~\bibnamefont {Uusitalo}}, \bibinfo {author} {\bibfnamefont
  {B.}~\bibnamefont {Cederwall}}, \bibinfo {author} {\bibfnamefont
  {A.}~\bibnamefont {Ertoprak}}, \bibinfo {author} {\bibfnamefont
  {H.}~\bibnamefont {Liu}}, \bibinfo {author} {\bibfnamefont {S.}~\bibnamefont
  {Guo}}, \bibinfo {author} {\bibfnamefont {M.~L.}\ \bibnamefont {Liu}},
  \bibinfo {author} {\bibfnamefont {J.~G.}\ \bibnamefont {Wang}}, \bibinfo
  {author} {\bibfnamefont {X.~H.}\ \bibnamefont {Zhou}}, \bibinfo {author}
  {\bibfnamefont {I.}~\bibnamefont {Kuti}}, \bibinfo {author} {\bibfnamefont
  {J.}~\bibnamefont {Tim\'ar}}, \bibinfo {author} {\bibfnamefont
  {A.}~\bibnamefont {Tucholski}}, \bibinfo {author} {\bibfnamefont
  {J.}~\bibnamefont {Srebrny}}, \ and\ \bibinfo {author} {\bibfnamefont
  {C.}~\bibnamefont {Andreoiu}},\ }\href {\doibase 10.1103/PhysRevC.100.024314}
  {\bibfield  {journal} {\bibinfo  {journal} {Phys. Rev. C}\ }\textbf {\bibinfo
  {volume} {100}},\ \bibinfo {pages} {024314} (\bibinfo {year}
  {2019})}\BibitemShut {NoStop}%
\bibitem [{\citenamefont {Wang}\ \emph {et~al.}(2019)\citenamefont {Wang},
  \citenamefont {Zhang}, \citenamefont {Zhao},\ and\ \citenamefont
  {Meng}}]{WANG2019454}%
  \BibitemOpen
  \bibfield  {author} {\bibinfo {author} {\bibfnamefont {Y.~Y.}\ \bibnamefont
  {Wang}}, \bibinfo {author} {\bibfnamefont {S.~Q.}\ \bibnamefont {Zhang}},
  \bibinfo {author} {\bibfnamefont {P.~W.}\ \bibnamefont {Zhao}}, \ and\
  \bibinfo {author} {\bibfnamefont {J.}~\bibnamefont {Meng}},\ }\href {\doibase
  https://doi.org/10.1016/j.physletb.2019.04.014} {\bibfield  {journal}
  {\bibinfo  {journal} {Phys. Lett. B}\ }\textbf {\bibinfo {volume} {792}},\
  \bibinfo {pages} {454 } (\bibinfo {year} {2019})}\BibitemShut {NoStop}%
\bibitem [{\citenamefont {Wang}\ \emph {et~al.}(2020)\citenamefont {Wang},
  \citenamefont {Wang},\ and\ \citenamefont {Meng}}]{wang2020}%
  \BibitemOpen
  \bibfield  {author} {\bibinfo {author} {\bibfnamefont {Y.~P.}\ \bibnamefont
  {Wang}}, \bibinfo {author} {\bibfnamefont {Y.~Y.}\ \bibnamefont {Wang}}, \
  and\ \bibinfo {author} {\bibfnamefont {J.}~\bibnamefont {Meng}},\ }\href@noop
  {} {\bibfield  {journal} {\bibinfo  {journal} {to be published}\ } (\bibinfo
  {year} {2020})}\BibitemShut {NoStop}%
\bibitem [{\citenamefont {Frauendorf}\ and\ \citenamefont
  {Meng}(1996)}]{Frauendorf1996Interpretation}%
  \BibitemOpen
  \bibfield  {author} {\bibinfo {author} {\bibfnamefont {S.}~\bibnamefont
  {Frauendorf}}\ and\ \bibinfo {author} {\bibfnamefont {J.}~\bibnamefont
  {Meng}},\ }\href {\doibase 10.1007/BF02769229} {\bibfield  {journal}
  {\bibinfo  {journal} {Z. Phys. A}\ }\textbf {\bibinfo {volume} {356}},\
  \bibinfo {pages} {263} (\bibinfo {year} {1996})}\BibitemShut {NoStop}%
\bibitem [{\citenamefont {Koike}\ \emph {et~al.}(2004)\citenamefont {Koike},
  \citenamefont {Starosta},\ and\ \citenamefont {Hamamoto}}]{Koike2004Chiral}%
  \BibitemOpen
  \bibfield  {author} {\bibinfo {author} {\bibfnamefont {T.}~\bibnamefont
  {Koike}}, \bibinfo {author} {\bibfnamefont {K.}~\bibnamefont {Starosta}}, \
  and\ \bibinfo {author} {\bibfnamefont {I.}~\bibnamefont {Hamamoto}},\ }\href
  {\doibase 10.1103/PhysRevLett.93.172502} {\bibfield  {journal} {\bibinfo
  {journal} {Phys. Rev. Lett.}\ }\textbf {\bibinfo {volume} {93}},\ \bibinfo
  {pages} {172502} (\bibinfo {year} {2004})}\BibitemShut {NoStop}%
\bibitem [{\citenamefont {Nilsson}\ \emph {et~al.}(1969)\citenamefont
  {Nilsson}, \citenamefont {Tsang}, \citenamefont {Sobiczewski}, \citenamefont
  {Szyma{\'n}ski}, \citenamefont {Wycech}, \citenamefont {Gustafson},
  \citenamefont {Lamm}, \citenamefont {M{\"o}ller},\ and\ \citenamefont
  {Nilsson}}]{nilsson1969nuclear}%
  \BibitemOpen
  \bibfield  {author} {\bibinfo {author} {\bibfnamefont {S.~G.}\ \bibnamefont
  {Nilsson}}, \bibinfo {author} {\bibfnamefont {C.~F.}\ \bibnamefont {Tsang}},
  \bibinfo {author} {\bibfnamefont {A.}~\bibnamefont {Sobiczewski}}, \bibinfo
  {author} {\bibfnamefont {Z.}~\bibnamefont {Szyma{\'n}ski}}, \bibinfo {author}
  {\bibfnamefont {S.}~\bibnamefont {Wycech}}, \bibinfo {author} {\bibfnamefont
  {C.}~\bibnamefont {Gustafson}}, \bibinfo {author} {\bibfnamefont {I.-L.}\
  \bibnamefont {Lamm}}, \bibinfo {author} {\bibfnamefont {P.}~\bibnamefont
  {M{\"o}ller}}, \ and\ \bibinfo {author} {\bibfnamefont {B.}~\bibnamefont
  {Nilsson}},\ }\href {\doibase 10.1016/0375-9474(69)90809-4} {\bibfield
  {journal} {\bibinfo  {journal} {Nucl. Phys. A}\ }\textbf {\bibinfo {volume}
  {131}},\ \bibinfo {pages} {1} (\bibinfo {year} {1969})}\BibitemShut {NoStop}%
\bibitem [{\citenamefont {Zhang}\ \emph {et~al.}(2007)\citenamefont {Zhang},
  \citenamefont {Qi}, \citenamefont {Wang},\ and\ \citenamefont
  {Meng}}]{Zhang2007Phys.Rev.C44307}%
  \BibitemOpen
  \bibfield  {author} {\bibinfo {author} {\bibfnamefont {S.~Q.}\ \bibnamefont
  {Zhang}}, \bibinfo {author} {\bibfnamefont {B.}~\bibnamefont {Qi}}, \bibinfo
  {author} {\bibfnamefont {S.~Y.}\ \bibnamefont {Wang}}, \ and\ \bibinfo
  {author} {\bibfnamefont {J.}~\bibnamefont {Meng}},\ }\href {\doibase
  10.1103/PhysRevC.75.044307} {\bibfield  {journal} {\bibinfo  {journal} {Phys.
  Rev. C}\ }\textbf {\bibinfo {volume} {75}},\ \bibinfo {pages} {044307}
  (\bibinfo {year} {2007})}\BibitemShut {NoStop}%
\bibitem [{\citenamefont {Qi}\ \emph {et~al.}(2009)\citenamefont {Qi},
  \citenamefont {Zhang}, \citenamefont {Meng}, \citenamefont {Wang},\ and\
  \citenamefont {Frauendorf}}]{Qi2009Chirality}%
  \BibitemOpen
  \bibfield  {author} {\bibinfo {author} {\bibfnamefont {B.}~\bibnamefont
  {Qi}}, \bibinfo {author} {\bibfnamefont {S.~Q.}\ \bibnamefont {Zhang}},
  \bibinfo {author} {\bibfnamefont {J.}~\bibnamefont {Meng}}, \bibinfo {author}
  {\bibfnamefont {S.~Y.}\ \bibnamefont {Wang}}, \ and\ \bibinfo {author}
  {\bibfnamefont {S.}~\bibnamefont {Frauendorf}},\ }\href {\doibase
  10.1016/j.physletb.2009.02.061} {\bibfield  {journal} {\bibinfo  {journal}
  {Phys. Lett. B}\ }\textbf {\bibinfo {volume} {675}},\ \bibinfo {pages} {175 }
  (\bibinfo {year} {2009})}\BibitemShut {NoStop}%
\bibitem [{\citenamefont {Wang}\ \emph {et~al.}(2007)\citenamefont {Wang},
  \citenamefont {Zhang}, \citenamefont {Qi},\ and\ \citenamefont
  {Meng}}]{Wang2007Phys.Rev.C24309}%
  \BibitemOpen
  \bibfield  {author} {\bibinfo {author} {\bibfnamefont {S.~Y.}\ \bibnamefont
  {Wang}}, \bibinfo {author} {\bibfnamefont {S.~Q.}\ \bibnamefont {Zhang}},
  \bibinfo {author} {\bibfnamefont {B.}~\bibnamefont {Qi}}, \ and\ \bibinfo
  {author} {\bibfnamefont {J.}~\bibnamefont {Meng}},\ }\href {\doibase
  10.1103/PhysRevC.75.024309} {\bibfield  {journal} {\bibinfo  {journal} {Phys.
  Rev. C}\ }\textbf {\bibinfo {volume} {75}},\ \bibinfo {pages} {024309}
  (\bibinfo {year} {2007})}\BibitemShut {NoStop}%
\bibitem [{\citenamefont {Chen}\ and\ \citenamefont
  {Meng}(2018)}]{Chen2018Phys.Rev.31303}%
  \BibitemOpen
  \bibfield  {author} {\bibinfo {author} {\bibfnamefont {Q.~B.}\ \bibnamefont
  {Chen}}\ and\ \bibinfo {author} {\bibfnamefont {J.}~\bibnamefont {Meng}},\
  }\href {\doibase 10.1103/PhysRevC.98.031303} {\bibfield  {journal} {\bibinfo
  {journal} {Phys. Rev. C}\ }\textbf {\bibinfo {volume} {98}},\ \bibinfo
  {pages} {031303(R)} (\bibinfo {year} {2018})}\BibitemShut {NoStop}%
\bibitem [{\citenamefont {Chen}\ \emph {et~al.}(2016)\citenamefont {Chen},
  \citenamefont {Zhang}, \citenamefont {Zhao}, \citenamefont {Jolos},\ and\
  \citenamefont {Meng}}]{Chen2016Two}%
  \BibitemOpen
  \bibfield  {author} {\bibinfo {author} {\bibfnamefont {Q.~B.}\ \bibnamefont
  {Chen}}, \bibinfo {author} {\bibfnamefont {S.~Q.}\ \bibnamefont {Zhang}},
  \bibinfo {author} {\bibfnamefont {P.~W.}\ \bibnamefont {Zhao}}, \bibinfo
  {author} {\bibfnamefont {R.~V.}\ \bibnamefont {Jolos}}, \ and\ \bibinfo
  {author} {\bibfnamefont {J.}~\bibnamefont {Meng}},\ }\href {\doibase
  10.1103/PhysRevC.94.044301} {\bibfield  {journal} {\bibinfo  {journal} {Phys.
  Rev. C}\ }\textbf {\bibinfo {volume} {94}},\ \bibinfo {pages} {044301}
  (\bibinfo {year} {2016})}\BibitemShut {NoStop}%
\bibitem [{\citenamefont {Chen}\ \emph {et~al.}(2017)\citenamefont {Chen},
  \citenamefont {Chen}, \citenamefont {Luo}, \citenamefont {Meng},\ and\
  \citenamefont {Zhang}}]{Chen2017Chiral}%
  \BibitemOpen
  \bibfield  {author} {\bibinfo {author} {\bibfnamefont {F.~Q.}\ \bibnamefont
  {Chen}}, \bibinfo {author} {\bibfnamefont {Q.~B.}\ \bibnamefont {Chen}},
  \bibinfo {author} {\bibfnamefont {Y.~A.}\ \bibnamefont {Luo}}, \bibinfo
  {author} {\bibfnamefont {J.}~\bibnamefont {Meng}}, \ and\ \bibinfo {author}
  {\bibfnamefont {S.~Q.}\ \bibnamefont {Zhang}},\ }\href {\doibase
  10.1103/PhysRevC.96.051303} {\bibfield  {journal} {\bibinfo  {journal} {Phys.
  Rev. C}\ }\textbf {\bibinfo {volume} {96}},\ \bibinfo {pages} {051303(R)}
  (\bibinfo {year} {2017})}\BibitemShut {NoStop}%
\end{thebibliography}
\end{document}